\documentclass[runningheads]{llncs}
\usepackage{graphicx}
\begin{document}
\title{An Unsupervised Approach to Ultrasound Elastography with End-to-end Strain Regularisation}
\titlerunning{An unsupervised approach to ultrasound elastography}

\author{R\'emi Delaunay\inst{1,2}\orcidID{0000-0002-0398-4995} \and
Yipeng Hu\inst{1}\orcidID{0000-0003-4902-0486} \and
Tom Vercauteren\inst{1,2}\orcidID{0000-0003-1794-0456}}


\authorrunning{R. Delaunay et al.}

\institute{Wellcome/EPSRC Centre for Interventional and Surgical Sciences, University College London, Gower Street, London WC1E~6BT, UK \and
School of Biomedical Engineering \& Imaging Sciences, King’s College London, Strand, London WC2R~2LS, UK \\
\email{remi.delaunay.17@ucl.ac.uk }}

\maketitle 

\begin{abstract}
Quasi-static ultrasound elastography (USE) is an imaging modality that consists of determining a measure of deformation (i.e. strain) of soft tissue in response to an applied mechanical force. The strain is generally determined by estimating the displacement between successive ultrasound frames acquired before and after applying  manual compression. The computational efficiency and accuracy of the displacement prediction, also known as time-delay estimation, are key challenges for real-time USE applications. In this paper, we present a novel deep-learning method for efficient time-delay estimation between ultrasound radio-frequency (RF) data. The proposed method consists of a convolutional neural network (CNN) that predicts a displacement field between a pair of pre- and post-compression ultrasound RF frames. The network is trained in an unsupervised way, by optimizing a similarity metric between the reference and compressed image. We also introduce a new regularization term that preserves displacement continuity by directly optimizing the strain smoothness. We validated the performance of our method by using both ultrasound simulation and \textit{in vivo} data on healthy volunteers. We also compared the performance of our method with a state-of-the-art method called OVERWIND~\cite{Mirzaei2019}. Average contrast-to-noise ratio (CNR) and signal-to-noise ratio (SNR) of our method in 30 simulation and 3 \textit{in vivo} image pairs are 7.70 and 6.95, 7 and 0.31, respectively. Our results suggest that our approach can effectively predict accurate strain images. The unsupervised aspect of our approach represents a great potential for the use of deep learning application for the analysis of clinical ultrasound data.
\end{abstract}

\section{Introduction}
Ultrasound elastography (USE) is an imaging technique that enables the characterization of tissue mechanical properties~\cite{Sigrist2017}. Since its introduction in 1991~\cite{Ophir1991a}, strain imaging has shown usefulness in diagnostic applications where pathological alterations induce modification of tissue stiffness, such as lesion detection in liver disease~\cite{Jeong2014} and tumour characterisation in the thyroid~\cite{Kwak2014}, breast~\cite{Hiltawsky2001} and prostate cancer~\cite{Correas2013}. This work focuses on quasi-static, free-hand strain elastography, where a time-varying axial compression is applied with an ultrasound transducer to the targeted tissue~\cite{Varghese2009}. The tissue mechanical behavior is then determined by mapping the relative deformation (i.e. strain) induced by manual compression (i.e. stress). 
\newline \indent 
Various methods have been proposed over the years to measure the strain. The main approach consists of determining the spatial displacement between a pair of radio-frequency (RF) ultrasound image data, acquired before and after applying an axial compression. The displacement estimation, also known as time-delay estimation, has been historically performed by maximizing a correlation function between local frame windows, either in the time or phase domain~\cite{Ophir1996,Shi2007,Chen2007,Varghese2000,Ara2013}. More recently, different approaches have added a regularization parameter to account for displacement discontinuity and improve displacement estimates~\cite{Kuzmin2015,Hashemi2017b}. Although these methods demonstrated the ability to make an accurate displacement estimation, current techniques are sensitive to noise and global decorrelation, i.e. the change of speckle appearance due to out-of-plane motion. Furthermore, real-time imaging is an important feature of elastography, and a trade-off is often made between the precision of standard approaches and their computational cost.
\newline \indent
The recent progress of learning-based methods in computer vision for optical flow estimation have inspired new approaches for strain elastography~\cite{Ilg2017a}. Those methods demonstrated the ability of neural networks to exploit the ultrasound high-frequency content and to robustly produce accurate displacement estimates~\cite{Kibria2018,Wu2018}. Previous networks have been trained using ultrasound simulation associated with ground truth displacement and strain labels. Accurate ground-truth images can be difficult to obtain for quasi-static elastography because the magnitude of applied stress is unknown. However, real-world ultrasound data often exhibits complex speckle patterns and echogenic features that can be quite challenging to replicate in ultrasound simulation. 
\newline \indent 
In this paper, we proposed an unsupervised method for time-delay estimation that allows a neural network to be trained directly on clinical data and predict tissue displacement. Unlike previous methods, our CNN training procedure is performed without ground truth labels. Instead, the network weights are optimized by minimizing a dissimilarity function between the pre-compression image and warped compressed image. We also introduce a new regularization term that preserves displacement continuity by directly optimizing the smoothness of the strain prediction. We validated our method by applying it on both real ultrasound data and simulations. The performance of our method was evaluated by comparing our displacement field and strain prediction with the ground truth labels and a gold standard USE technique~\cite{Mirzaei2019}. To the best of our knowledge, this is the first unsupervised deep-learning method applied to strain imaging.

\section{Methods}
\label{methods}
\subsection{Problem Statement}
\label{problem}
\begin{figure} \begin{center}
  \includegraphics[width=0.80\textwidth]{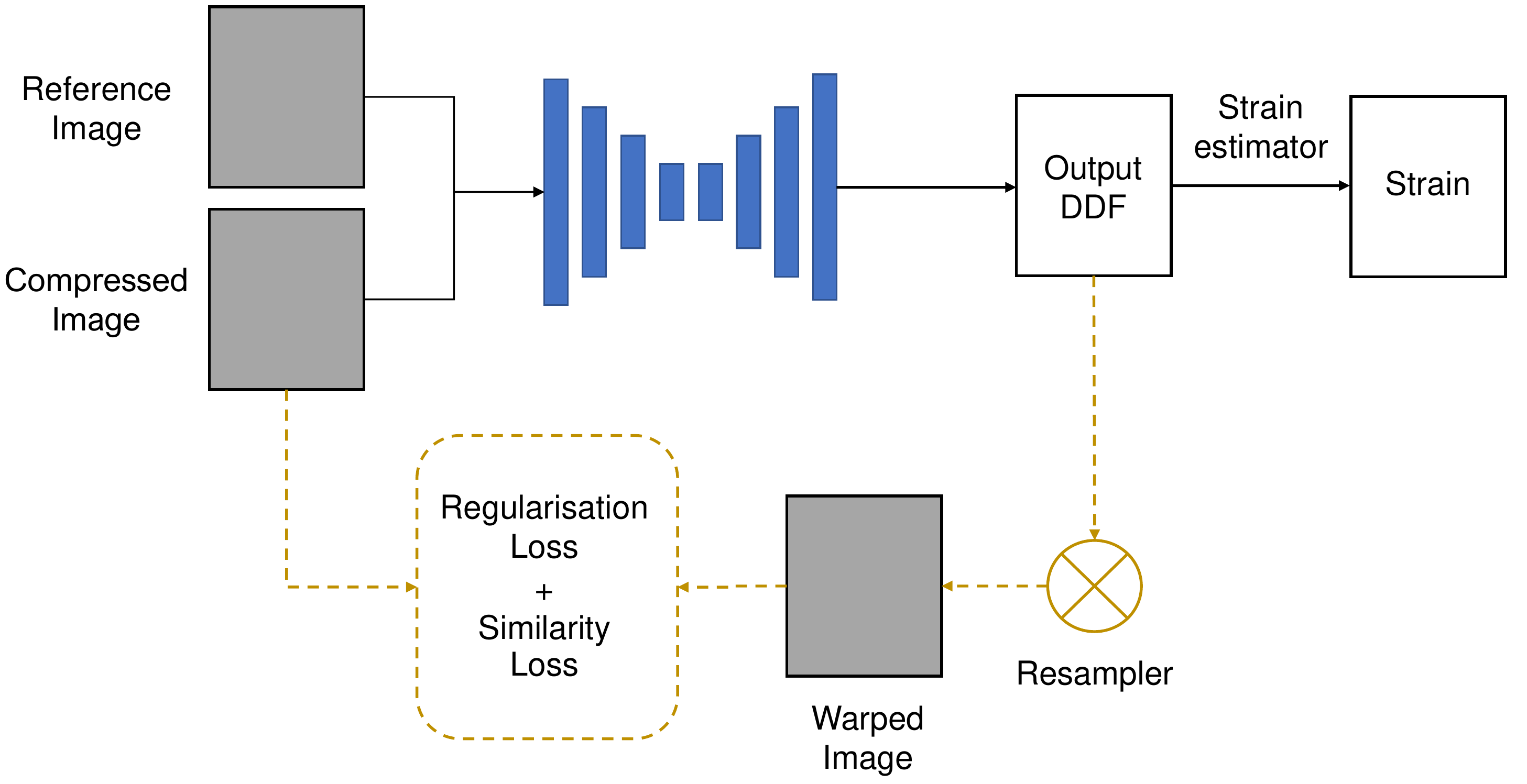}
\caption{Overview of the method.}
\label{fig:1} 
\end{center}
\end{figure}

Our method follows an approach similar to a standard image registration framework, which aims to find a spatial transformation that maps a moving image into the space of a reference image. In the case of a non-rigid image registration solution, this transformation can be represented as a dense displacement field (DDF). This transformation is generally optimized in an iterative manner, by maximizing an objective function that measures the similarity between the warped moving image and the reference image, and a regularization parameter to ensure displacement continuity. From a learning-based approach perspective, the image mapping is predicted by a neural network instead of being directly optimized. An overview of the method is presented in Fig.~\ref{fig:1}.

\subsection{Displacement Estimation}
\label{timedelay}
Our network takes as input a pair of pre- and post-compression 2D RF frames, here named $Pre$ and $Post$, and predicts a DDF $u$. The network parameters are learned by minimizing a dissimilarity metric $L_{sim}$ between pairs of 2D RF ultrasound data. Our training loss function also includes a regularization term, $L_{reg}$, acting on the predicted displacement field $u$ and associated with a weighting hyper-parameter $\alpha$, to ensure balance between the likelihood and smoothness of the predicted transformation. The optimization problem can be written as:
\begin{equation}
\hat{\theta} = \arg\min_\theta \ [ L_{sim}(Pre,Post;u)\ +\ \alpha \ .\ L_{reg}(u)]
\end{equation}
where $\theta$ represents the network parameters that are optimized through stochastic gradient descent. 

The dissimilarity metric corresponds to a negative local normalized cross-correlation (NCC) which average the NCC score between sliding windows sampled from the pre-compression image and the transformed post-compression image, resampled with the predicted displacement field. The NCC between two local image windows, $W_{pre}$ and $W_{post}$, with $i,j$ pixel components can be written as:
\begin{equation}
NCC=\frac{1}{N}\sum_{i,j}\frac{\left[W_{pre}(i,j) - \mu_{W_{pre}} \right ] \times  \left [W_{post}(i,j) - \mu_{W_{post}} \right ]}{\sigma_{W_{pre}} \times \sigma_{W_{post}}}
\end{equation}
where $N$ is the number of pixels ($i,j$) and $\mu$ and $\sigma$ correspond to the mean and standard deviation of the images, respectively.

The regularisation term consists of the L1-norm of the second spatial derivatives of the predicted displacement field $u$. Given that the strain modulus corresponds to the displacement gradient, minimizing its second derivative allows to enforce the strain map smoothness. The regularisation term can be written as:
\begin{equation}
L_{reg} = \sum_{i,j}\ (| \ {\partial_x^2 u_{i,j}} \ | + | \ {\partial_y^2 u_{i,j}} \ |)
\end{equation}
where $\partial_x^2$ and $\partial_y^2$ are the second partial derivatives in axial and lateral directions, respectively.

\subsection{Implementation}
\subsubsection{Network}

The architecture of our network was presented in a method for medical image registration~\cite{Hu}, and consists in an encoder-decoder CNN with skip connections. The encoder part is composed of four down-sampling blocks, which capture the hierarchical features necessary to establish correspondence between the image pair. Each down-sampling block consists of two convolutional layers with a residual network shortcut, batch normalization, and leaky rectified linear unit (Leaky ReLU). Symmetrically, the decoder part is composed of four up-sampling blocks that consists of an additive up-sampling layer summed over a transpose convolutional layer. Finally, each up-sampling block outputs a displacement field that is convoluted and resized to the input size, then summed to output the predicted displacement field. The network was implemented in TensorFlow using NiftyNet~\cite{Gibson2018a}. It was trained by using the Adam optimiser, starting at a learning rate of 10-3, with a minibatch of 4. The regularisation weight was set to $\alpha=2$ to ensure displacement continuity. 

\subsubsection{Strain Estimation}
In USE, the strain estimates are obtained by computing the displacement field gradient. However, direct differentiation of the displacement field is rarely used because gradient operations generate a significant amount of noise in the resulting strain map. We used the least-squares strain estimator (LSQSE) to improve the elastogram signal-to-noise ratio~\cite{Kallel1997}. The strain estimation was also implemented in TensorFlow to facilitate efficient parallel computing. In inference, the strain map  prediction rate reach a total of 13 images per second on a 12GB NVIDIA GTX-1080ti GPU.

\section{Experiments}
\label{experiments}
The performance of our method was evaluated on ultrasound simulations and \textit{in vivo} data. We compared our results with a state-of-the-art strain elastography method called “tOtal Variation Regularization and WINDow-based time delay estimation” (OVERWIND)~\cite{Mirzaei2019}. The OVERWIND results were obtained by using the publicly available MATLAB implementation, and default parameters were chosen for the simulation experiment. The regularisation coefficients were manually tuned for the \textit{in vivo} data to globally maximize the NCC and Contrast-to-Noise Ratio (CNR) scores of the three cases. The results on simulation were also compared with strain estimates obtained by training our network with a supervised loss function. The supervised loss function was used before for time-delay estimation~\cite{Wu2018}, and corresponded to the mean absolute difference (MAE) between the network prediction and the ground truth labels. The quality of the strain estimates were assessed with the CNR and Signal-to-Noise Ratio (SNR), which are metrics commonly used in USE~\cite{Wang2018,Mirzaei2019}.

\subsubsection{Simulation Dataset}

The Field-II software~\cite{Jensen1992,Jensen1996} was used to generate the ultrasound images. Each simulation consisted in a 3D rectangle of size 38x40 mm, containing a cylindrical-shaped inclusion with a randomly assigned diameter (from 8 to 12 mm) and position. The speckle pattern typically observed in ultrasound imaging was obtained by randomly assigning a total of 400,000 scatterers across each digital phantom. The axial compression was assigned randomly to each phantom and represented between 0.5\% and 4\% of the phantom total length. The Young's modulus of the inclusion was set to different values (i.e. 8, 15, 45 and 75 kPa) while the background was fixed to 25 kPa. Tissue displacements were estimated by finite element method using the Partial Differential Equation Toolbox from MATLAB, and were used to interpolate the scatterers position. A total of 192 RF lines were simulated for each image, with probe central and sampling frequency of 7 and 40 MHz, respectively. The background scatterers were associated with a random intensity value from a Gaussian distribution, to mimic homogeneous tissue. To increase the network's robustness to noise and image intensities, the inclusion scatterers intensities were either assigned to 0 or similar to the background. Moreover, white Gaussian noise with random signal power values, from 5 to 20 dBw, was added on each image to increase robustness to noise. Finally, 160 ultrasound image pairs were simulated, where 100 were used for training and 30 for validation and testing, respectively.

\subsubsection{\textit{In Vivo} Dataset}
We created our own \textit{in vivo} dataset by collecting images of the arms in three human volunteers. The RF data was acquired  from a Cicada 128PX system equipped with a 10 MHz linear probe from Cephasonics (Cephasonics Inc., USA). The images were reconstructed using the delay-and-sum beamformer from SUPRA~\cite{Gobl2017}. Experimental protocol consisted in acquiring sequences of images while slowly applying an axial compression on the volunteer's arm with the handheld ultrasound probe. The \textit{in vivo} dataset included 1300 image pairs for training and 300 pairs for validation and testing. The three image pairs presented in the results section were taken from the testing partition and exhibit one or several blood vessels located in the arm.

\subsection{Results}
\begin{figure}
\begin{center}
\includegraphics[width=0.7\textwidth]{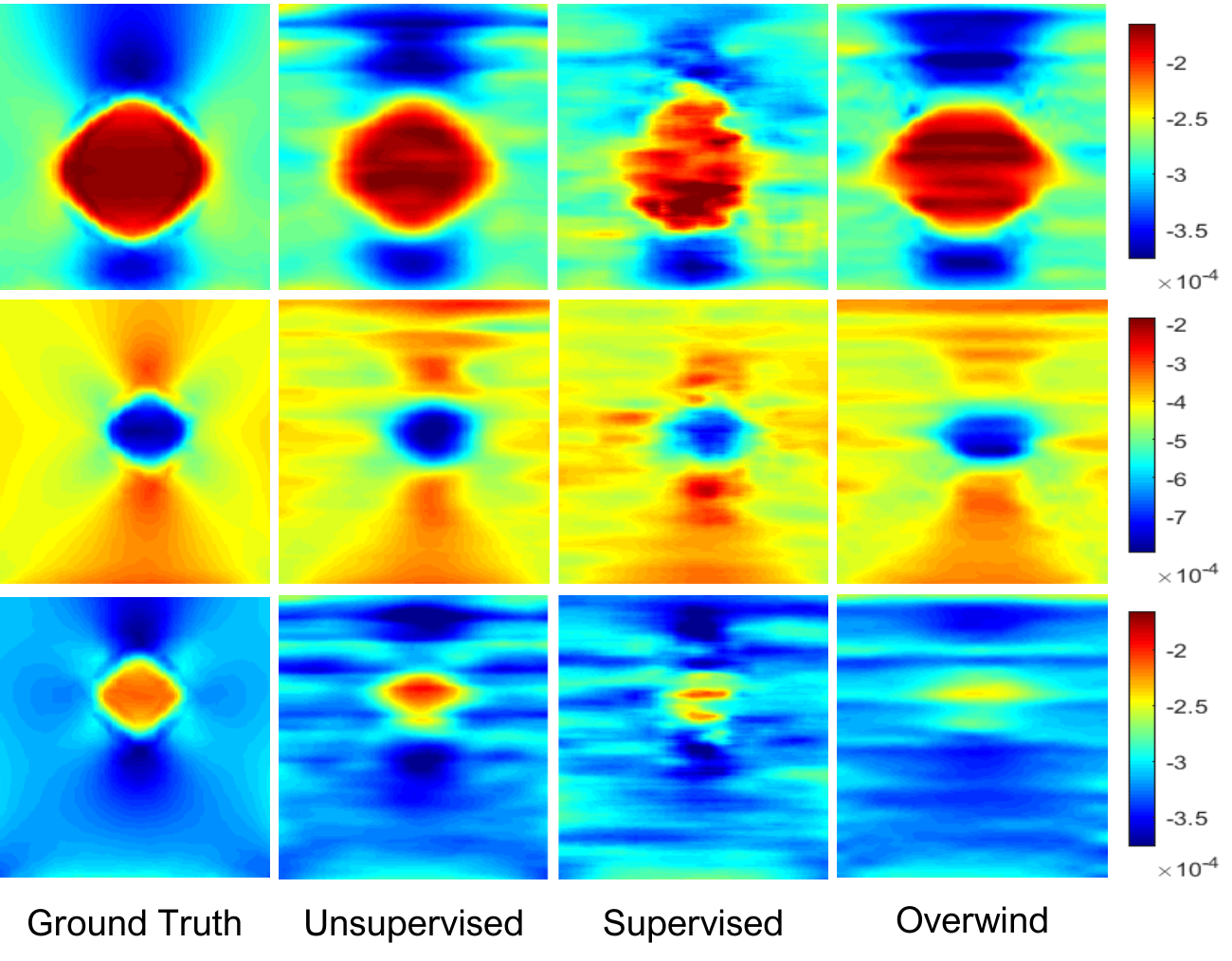}
\caption{Comparison of three axial strain fields computed from ultrasound simulations by finite element method (i.e., ground truth), OVERWIND, the unsupervised and supervised network. OVERWIND's regularisation parameters : $\alpha1=\beta1=20$ and $\alpha2=\beta2=8$.}
\label{fig:2}
\end{center}
\end{figure}

Figure \ref{fig:2} shows axial strain images estimated from three simulated image pairs and obtained with the finite element method, OVERWIND and both supervised and unsupervised models. The averaged CNR and SNR values of the entire testing dataset (i.e., 30 image pairs) are displayed in table \ref{tab:1}.

\begin{table}[!htb]
\begin{center}
\caption{Mean and standard deviation of SNR and CNR for the strain images of the simulation testing dataset obtained with finite element method, OVERWIND, the unsupervised and supervised network.}
\label{tab:1}
\begin{tabular}{ccccc}
\hline
    & \begin{tabular}[c]{@{}c@{}}Ground truth\\ (mean std)\end{tabular} & \begin{tabular}[c]{@{}c@{}}Unsupervised\\ (mean std)\end{tabular} & \begin{tabular}[c]{@{}c@{}}Supervised\\ (mean std)\end{tabular} & \begin{tabular}[c]{@{}c@{}}OVERWIND\\ (mean std)\end{tabular} \\ \hline
SNR & 7.51 (2.61) & 6.95 (2.54) & 7.79 (2.01) & 9.31 (3.51) \\ 
CNR & 9.15 (2.73) & 7.70 ( 3.8) & 4.22 (2.08) & 6.33 (3.6) \\ \hline
\end{tabular}
\end{center}
\end{table}

\begin{figure}
\begin{center}
\includegraphics[width=0.7\textwidth]{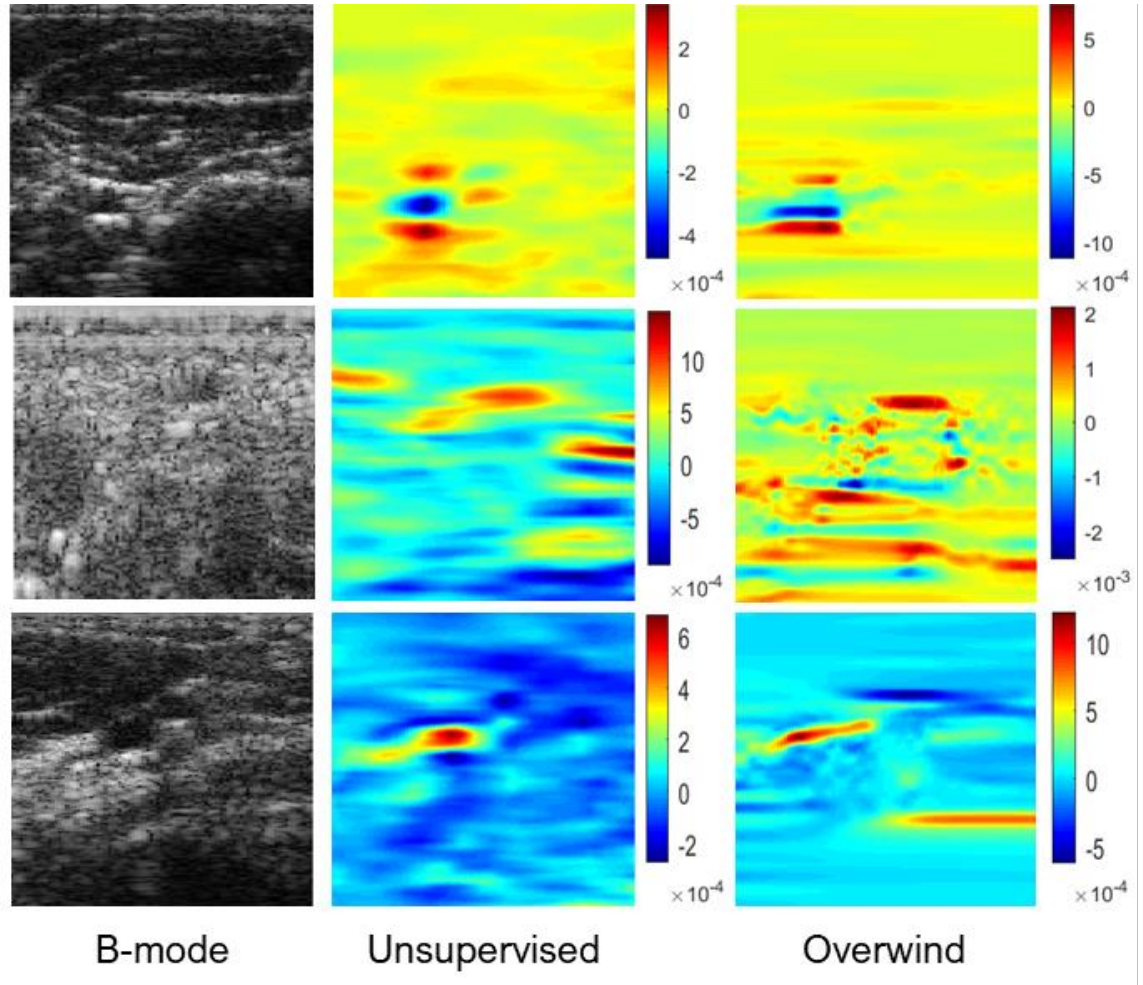}
\caption{Comparison of axial strains estimated by our method and OVERWIND on three image pairs from the \textit{in vivo} testing dataset . OVERWIND's regularisation parameters : $\alpha1 = \beta1=0.2$ and $  \alpha2=\beta2=0.05$.}
\label{fig:3}
\end{center}
\end{figure}

Figure \ref{fig:3} shows axial strains estimated from three different \textit{in vivo} image pairs. CNR and SNR values of each case can be found in table \ref{tab:2}. The local NCC values between the post-compression and resampled pre-compression images are also displayed, to indicate the quality of the predicted displacement fields. The mean of the three CNR values are 7 and 8.43 for our unsupervised model and OVERWIND, respectively. The average SNR values are 0.31 and 0.7 for our method and OVERWIND, respectively. Displacement field estimates from both the simulation and \textit{in vivo} experiments are available in the supplementary material.

\begin{table}[!htb]
\begin{center}
\caption{SNR and CNR of strain estimates from our unsupervised method and OVERWIND, and local NCC scores for the three image pair results from the \textit{in vivo} dataset.}
\label{tab:2}       
\begin{tabular}{ccccccc}
\hline

{} & \multicolumn{3}{c}{Unsupervised} & \multicolumn{3}{c}{OVERWIND} \\ 
                  & Case 1    & Case 2    & Case 3    & Case 1   & Case 2   & Case 3  \\ \hline
SNR               & 0.14      & 0.07      & 0.1       & 0.12     & 0.35     & 0.23    \\ 
CNR               & 6.22      & 5.2       & 4.5       & 4.9      & 3.15     & 2.23    \\ 
LNCC               & 0.87      & 0.81      & 0.94      & 0.90     & 0.81     & 0.93    \\ \hline
\end{tabular}
\end{center}
\end{table}

\section{Discussion}
In this work, we presented a new deep-learning approach for the estimation of the displacement and strain maps between a pair of ultrasound RF data undergoing an axial compression. We validated our method on both ultrasound simulation and \textit{in vivo} data . Our method is completely unsupervised and ground truth images collected from finite element analysis were only used to assess the performance of our method. 

Our results on ultrasound simulation indicate that our method predicts strain estimates with a significantly better CNR compared to the supervised network, with 7.70 and 4.20 respectively. This suggests that our training loss function, which includes a term that penalizes the strain smoothness, improves the strain contrast. Our experiments showed comparable results to OVERWIND, a state-of-the-art method which has, in terms of CNR, already outperformed a previous classical approach~\cite{Hashemi2017b} by 27.26\%, 144.05\%, and 49.90\% on average in simulation, phantom, and in-vivo data, respectively, as reported in~\cite{Mirzaei2019}. In addition, our method is fully automatic in inference while OVERWIND strain estimation relies on the correct adjustment of its regularisation parameters.

Finally, the OVERWIND real-time performance had not been quantitatively reported while our approach reached a strain prediction rate of about 13 frames per second on a 12GB NVIDIA GTX-1080ti GPU. The real-time inference of our network and its ability to be trained without ground truth labels represents a great potential for the use of learning-based methods in ultrasound strain elastography.

\subsubsection{Acknowledgement}
This work was supported by the EPSRC [NS/A000049/1], [NS/A000050/1], [EP/L016478/1] and the Wellcome Trust [203148/Z/16/Z], [203145/Z/16/Z]. Tom Vercauteren is supported by a Medtronic/RAEng Research Chair [RCSRF1819/7/34].

\bibliographystyle{splncs04}
\bibliography{paper2405}

\end{document}